\pdfoutput=1
\documentclass[10pt,conference]{IEEEtran}
\usepackage{latexsym,url}
\usepackage{amsmath,amssymb,graphicx}
\usepackage{citesort}
\usepackage{epsf}

\long\def\comment#1{}

\newfont{\bbb}{msbm10 scaled 700}

\newfont{\bb}{msbm10 scaled 1100}

\renewcommand{\Re}{{\rm Re}}
\renewcommand{\Im}{{\rm Im}}

\usepackage{graphics} 
\usepackage{epsfig} 
\usepackage{amsmath} 
\usepackage{amssymb}  
\usepackage{dsfont}

\newtheorem{thm}{Theorem}

\long\def\symbolfootnote[#1]#2{\begingroup%
\def\thefootnote{\fnsymbol{footnote}}\footnote[#1]{#2}\endgroup}

\begin{document}
\title{\LARGE \bf Power Allocation Strategies and  Lattice Based Coding schemes for Bi-directional relaying
}
\author{ Makesh Pravin Wilson and  Krishna Narayanan\\
 Department of Electrical
and Computer Engineering,\\
        Texas A\&M University, College Station, TX 77843, USA.\\
        {\tt\small {makeshpravin@neo.tamu.edu,krn}@ece.tamu.edu}}
\maketitle



%

\begin{abstract}
We consider a communication system where two transmitters wish to
exchange information through a half-duplex relay in the middle. The
channels between the transmitters and the relay have asymmetric
channel gains. More specifically, the channels are assumed to be
synchronized with  complex inputs and complex fading coefficients
with an average power constraint on the inputs to the channels. The
noise at the receivers have the same power spectral density and are
assumed to be white and Gaussian. We restrict our attention to
transmission schemes where information from the two nodes are
simultaneously sent to the relay during a medium access phase
followed by a broadcast phase where the relay broadcasts information
to both the nodes. An upper bound on the capacity for the two phase
protocol under a sum power constraint on the transmit power from all
the nodes is obtained as a solution to a convex optimization
problem. We show that a scheme using channel inversion with lattice
decoding can obtain a rate a small constant $0.09$ bits from the
upper bound at high signal-to-noise ratios. Numerical results show
that the proposed scheme can perform very close to the upper bound.
\end{abstract}

\section{Introduction}\label{sec:priorwork}
In our previous work \cite{narayanan07}, we studied the
bi-directional relay problem where two users exchange information
with each other through a relay in the middle. Therein we assumed
same channel gains between the nodes and the relay. In this work we
look at a more practical scenario with asymmetric channel gains.
There has been recent work on the bi-directional relay problem with
asymmetric channel gains. In \cite{SangJoonKim08}, several schemes
including compress and forward, amplify and forward, decode and
forward and mixed forward have been suggested. In our work, we show
the benefit of using nested lattices for this problem. There has
been some work on using nested lattices for asymmetric channel gains
in \cite{baik08,nam08}. In \cite{nam08}, a scheme that is optimal at
high signal-to-noise ratios (snr) is given for a given realization
of the asymmetric channel.

In this paper, we consider a scenario of a fading channel with $L$
channel realizations and assume that channel gains are known to all
the nodes. The transmit power is adapted as a function of the
channel gains. We show that an upper bound to the achievable
exchange capacity (defined later) can be obtained as the solution to
a convex optimization problem. We also show that symmetric nested
lattices with an appropriate power allocation policy can perform
close to the upper bound at high SNRs.
 \symbolfootnote[0]{This work
was supported by the National Science Foundation under grant
CCR-0515296.}

\section{System Model and Problem Statement}\label{sec:System model}
 \begin{figure}[thpb]
      \centering
      \includegraphics[scale=0.3,angle =90]{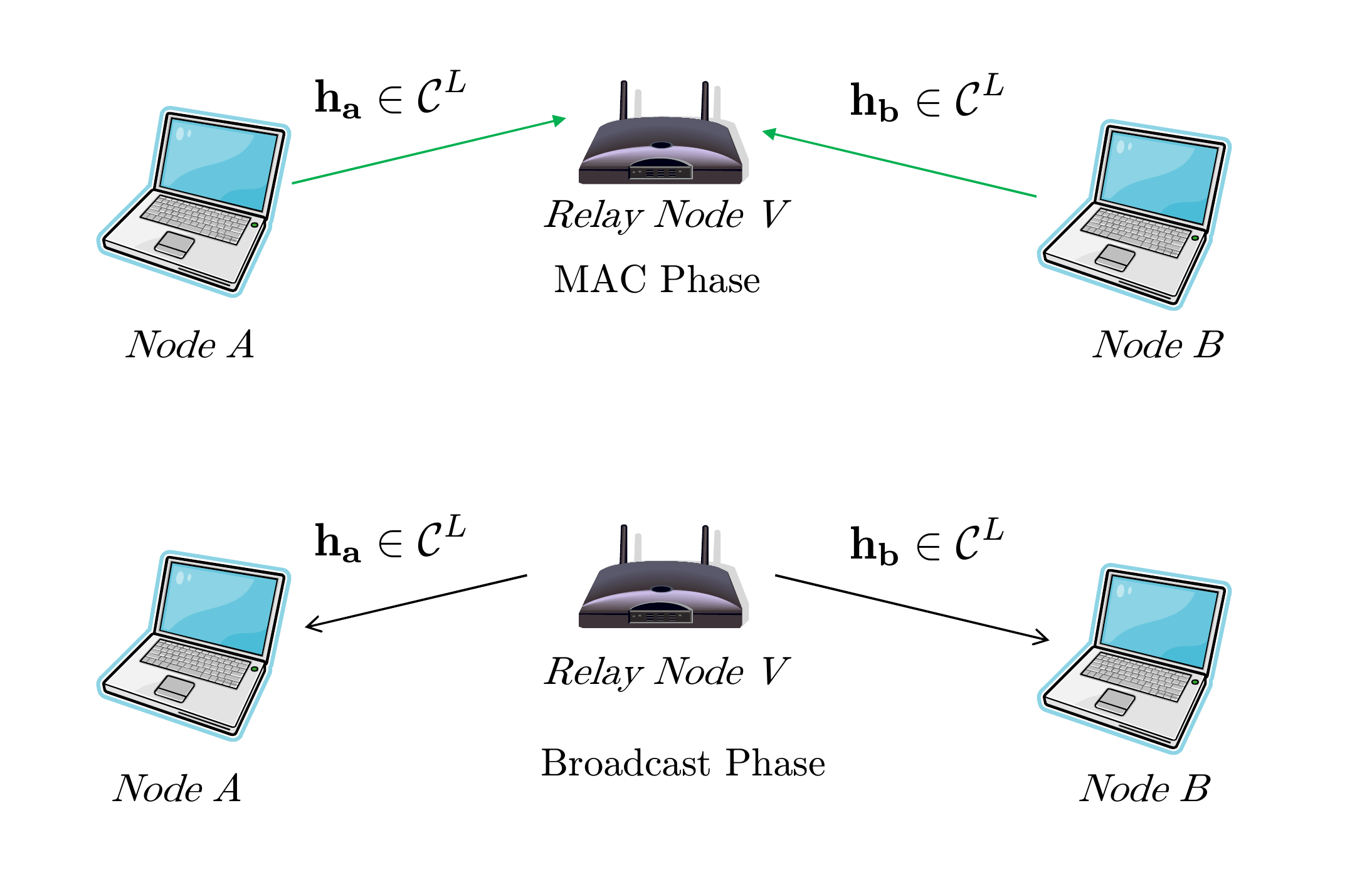}
      \caption{Problem setup with fading links $h_a$ and $h_b$}
      \label{fig:fading system model}
   \end{figure}
 We study a simple 3-node linear
Gaussian network, with asymmetric channel gains as shown in
Fig.~\ref{fig:fading system model}. Node $A$ and node $B$ wish to
exchange information between each other through the relay node $R$.
However, nodes $A$ and $B$ cannot communicate with each other
directly. The nodes are assumed to be half-duplex, i.e. a node can
not transmit and listen at the same time. We use the block fading
model to model the channels between the nodes and the relay. The
channel gains between the nodes and the relays remain constant over
a coherence time interval. It is assumed that the transmission
happens over $L$ such coherence time intervals. Each coherence
interval corresponds to $N$ uses of the channel. Hence, effectively
$NL$ uses of the channel are available for communication. For the
channel between the node $A$ and the relay $R$, each of  the
coefficients of the $L$ length vector $\mathbf{h_{ar}} \in
\mathbb{C}^L$, represent the channel gain for each coherence time
interval. Similarly the channel between the node $B$ and the relay
$R$, the relay $R$ and node $A$ and the relay $R$ and node $B$ are
captured by the channel gain vectors $\mathbf{h_{br}}$,
$\mathbf{h_{ra}}$ and $\mathbf{h_{rb}} \in \mathbb{C}^L$
respectively(vectors are denoted by bold face letters such as
$\mathbf{v}$ throughout the paper). The channels are assumed to be
reciprocal, i.e., $\mathbf{h_{ar}} = \mathbf{h_{ra}} :=
\mathbf{h_a}$ and $\mathbf{h_{br}} = \mathbf{h_{rb}} :=
\mathbf{h_b}$.

  Let $\mathbf{u}_A \in
\{0,1\}^{k_aL}$ and $\mathbf{u}_B \in \{0,1\}^{k_bL}$ be the
information vectors at nodes $A$ and $B$. We assume a protocol where
the communication takes place in two phases at  each coherence time
interval $i$ ( $i \in \{1 ,2 \hdots L  \}$). The phases are the
multiple access (MAC) phase and the broadcast phase. $\Delta \in [0,
1]$ is the fraction of channel uses for which the MAC phase is used
and $(1 - \Delta)$ is the fraction of channel uses for which  the
broadcast phase is used. It is assumed that communication in the MAC
and broadcast phases are orthogonal. For example, this could be in
two separate frequency bands (or in two different time slots) and
 hence the MAC phase and broadcast phase do not interfere with each other.

\paragraph{MAC phase} During the MAC phase of each coherence time interval $i$,
 nodes $A$ and $B$ transmit while the relay node listens.
  $\mathbf{x_{ai}} \in
\mathbb{C}^{\Delta N} $ and $\mathbf{x_{bi}} \in \mathbb{C}^{\Delta
N} $ are the transmitted vectors at nodes $A$ and $B$, respectively.
The MAC phase takes place in $\Delta N$ uses of the complex additive
white Gaussian Channel (AWGN) channel. Further it is assumed that
the two transmissions are perfectly synchronized. Hence the received
signal at the relay $\mathbf{y_{ri}}  \in \mathbb{C}^{\Delta N}$, is
given by
\[
\mathbf{y_{ri}} = h_{ai} \mathbf{x_{ai}} + h_{bi} \mathbf{x_{bi}} +
\mathbf{z_{ri}}
\]
where the components of $\mathbf{z_{ri}} \in \mathbb{C}^{\Delta N}$
are independent identically distributed (i.i.d) complex, circularly
symmetric Gaussian random variables with zero mean and unit
variance. The average transmit power at the nodes $A$ and $B$ in the
$i^{th}$ coherence time interval is given as $E[||X_{ai}||^2] =
P_{ai}$ and $E[||X_{bi}||^2] = P_{bi}$.

\paragraph{Broadcast phase} During the broadcast phase, the relay node transmits
$\mathbf{x_{ri}} \in \mathbb{C}^{(1-\Delta)N}$ in the $i^{th}$
coherence time interval to both nodes $A$ and $B$. The nodes A and B
receive $\mathbf{y_{ai}}$ and $\mathbf{y_{bi}}$, respectively where
\begin{eqnarray}
\mathbf{y_{ai}} &=& h_{ai} \mathbf{x_{ai}} + \mathbf{z_{ai}}
\nonumber \\
\mathbf{y_{bi}} &=& h_{bi} \mathbf{x_{bi}} + \mathbf{z_{bi}}
\nonumber
\end{eqnarray}
 The average transmit power
at the relay node during the $i^{th}$ coherence time interval is
given by $P_{ri}$, and the receiver noise at the two nodes is
complex Gaussian with zero mean and unit variance.

 Further, it is assumed that there is a total sum power
constraint over all the nodes. Since the MAC phase is used during
the fraction $\Delta$ and the broadcast phase is used during the
fraction $(1 - \Delta)$ of the available time slots, the total power
constraint is expressed as
\[
\frac{\Delta}{L} \sum_{i=1}^{L} P_{ai} + \frac{\Delta}{L}
\sum_{i=1}^{L} P_{bi} + \frac{(1 - \Delta)}{L} \sum_{i = 1}^{L}
P_{ri} \leq  P.
\]
We are interested in power allocation strategies and good
encoding/decoding schemes that maximize the amount of information
(maximize $k_a/(LN) + k_b/(LN)$) that can be exchanged reliably
(such that the probability of error can be made arbitrarily small in
the limit of $N \rightarrow \infty$). We refer to the maximum value
of $k_a/(LN) + k_b/(LN)$ that can be reliably exchanged with a given
scheme as the exchange rate for that scheme. The exchange capacity
is then the supremum of all such rates over the encoding schemes.

\section{Main Results and comments}
\label{sec:mainresults} We only consider the case when $\Delta =
0.5$, or the MAC and broadcast phase each use the channel half the
time. For this case, the main results in this paper are

\begin{itemize}

    \item An upper bound on the exchange rate is setup as a convex
    optimization problem.

    \item A scheme is proposed which at high snrs is away from the upper bound by at most 0.09 bits(see Theorem 3).
     The scheme uses nested lattice encoding and the transmit power is chosen to be inversely proportional to the channel gains.
\end{itemize}

\section{Upper Bound for the two phase
protocol}\label{sec:upperbound}

     We can easily obtain an upper bound for the two phase protocol
     using cut-set arguments and as in \cite{SangJoonKim08}. In our
     problem model the channel remains constant in each coherence
     time interval. Hence, the channel over $L$ such coherence time
     intervals can be modeled as a set of $L$ parallel channels.
     At each coherence time interval $i$, the maximum information rate that can be transmitted from node
     $A$ to node $B$ is bounded by the minimum of the information capacity from
     node $A$ to relay node $R$, and relay node $R$ to node $B$. This can be expressed as,
     $\min \{ \Delta C(|h_{ai}|^2P_{ai}), (1-\Delta) C(|h_{bi}|^2P_{ri})
     \}$, where $C(x):= \log (1 + x)$. Here $\Delta$ is the fraction
     of time node $A$ transmits and $(1 -\Delta)$ is the fraction of
     time the relay node transmits.
      Similarly
     the rate that can be transmitted from node $B$ to node $A$ is bounded by the
     minimum of the information capacity from node $B$ to relay $R$ and from relay $R$ to node
     $A$, which can be expressed as $\min \{ \Delta C(|h_{bi}|^2P_{bi}), (1-\Delta) C(|h_{ai}|^2P_{ri}) \} $.
      Hence the total rate that can be transmitted over $L$ such coherence time intervals
     can be expressed as the sum of the rates at each coherence time interval. Combining the above, the  upper bound on the
     exchange capacity can be expressed as the solution of an optimization problem given by,
       \begin{eqnarray}
         \mbox{maximize     }  && \frac{1}{L} \sum_{i=1}^{L}  \min \{ \Delta C(|h_{ai}|^2P_{ai}), (1-\Delta) C(|h_{bi}|^2P_{ri})
         \}\nonumber\\ &&               +  \frac{1}{L} \sum_{i=1}^{L}\min \{ \Delta C(|h_{bi}|^2P_{bi}), (1-\Delta) C(|h_{ai}|^2P_{ri}) \} \nonumber  \\
         \mbox{subject to     } && \frac{\Delta}{L} \sum_{i=1}^{L} P_{ai} + \frac{\Delta}{L} \sum_{i=1}^{L} P_{bi} + \frac{(1 -
\Delta)}{L} \sum_{i = 1}^{L} P_{ri} \leq P\nonumber \\
                                && P_{ai}, P_{bi}, P_{ri} \geq 0, i
                                \in {1,2 \hdots L.}
                                \label{eqn:upperbound}
       \end{eqnarray}
       Since $\Delta$ is fixed, we can see that the objective function given above is concave
 and also the constraints form a convex set. Hence convex
 optimization techniques can be easily applied to this setup to get
 the optimal solution for the above convex problem.

Though the problem is convex, it is tough to get analytical results
for general snr. However to develop some intuition and for ease of
analysis, we obtain the power allocation strategy by solving the
optimization problem under the high snr approximation, i.e. $C(x) :=
\log(x)$. We make this precise in the following theorem.
 \begin{figure}[thpb]
      \centering
      \includegraphics[scale=0.4]{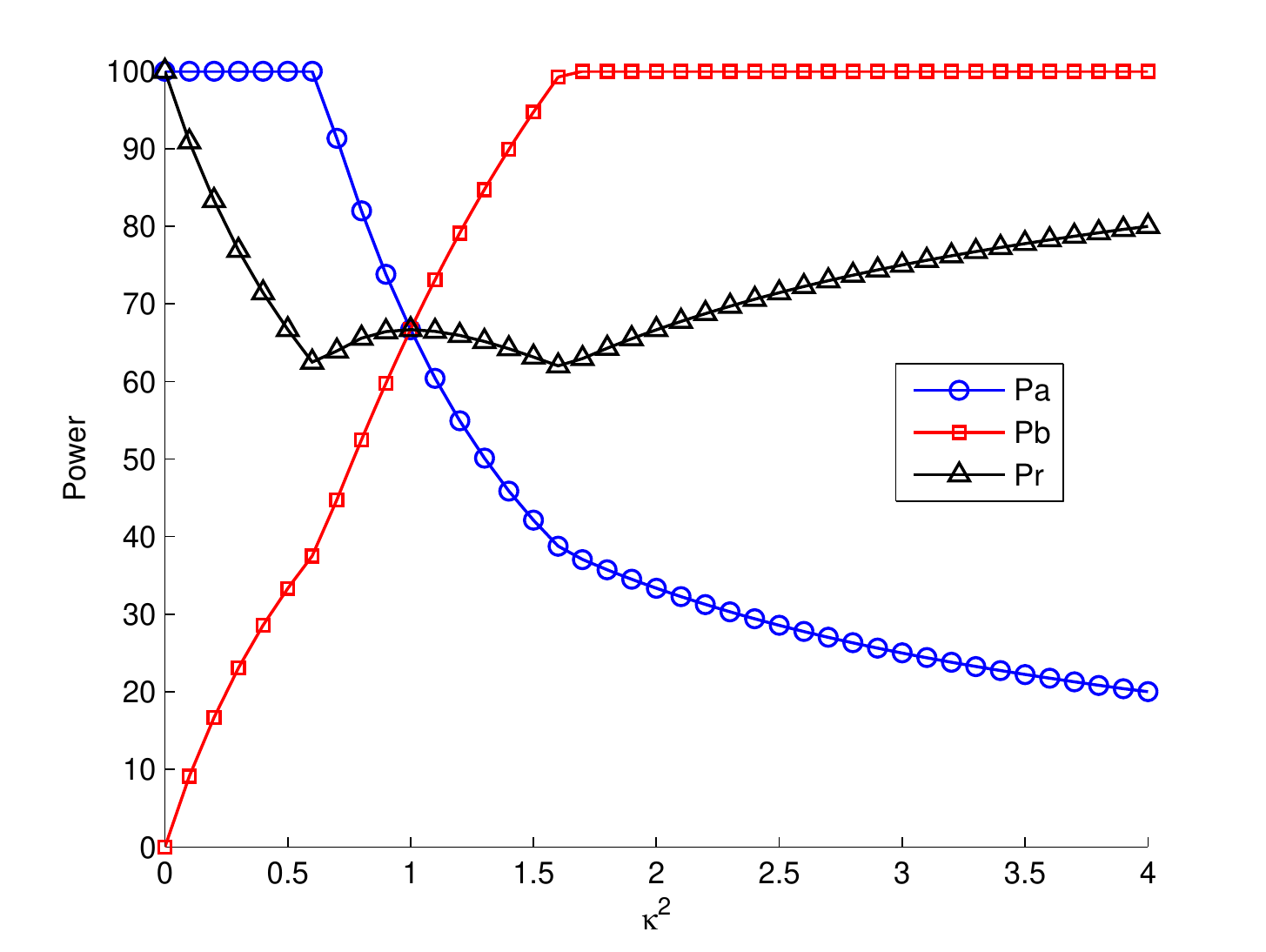}
      \caption{Power allocation as a function of $\kappa^2$ for the upper bound}
      \label{fig:upper_bd_power_profile}
   \end{figure}
\begin{thm}
For finite $L$ and $\Delta = 0.5$, and under the high snr
approximation, the optimal power allocation for the upper bound on
the achievable rate for different ranges of $\kappa_i^2 :=
|h_{ai}|^2/|h_{bi}|^2$ is given by,
\paragraph*{Case 1} $ 0 < \kappa_i^2 <= \left.\frac{-1 +
    \sqrt{5}}{2}\right.,$
\begin{equation*}P_{ai} = P  \,\,\,\,\,\,  P_{bi} = P \frac{\kappa_i^2}{1 + \kappa_i^2}
\,\,\,\,\,\, P_{ri} = P \frac{1}{1 + \kappa_i^2}
\end{equation*}
\paragraph*{Case 2} $ \left.\frac{-1 +
    \sqrt{5}}{2}\right. < \kappa_i^2 <= \left.\frac{1 +
    \sqrt{5}}{2}\right.,$
\begin{equation*}\begin{split}P_{ai} =
P \frac{2}{1 + \kappa_i^2 + \kappa_i^4}  \,\,\,\,\,\,  P_{bi} = P
\frac{2 \kappa_i^4}{1 + \kappa_i^2 + \kappa_i^4} \,\,\,\,\,\,
\\P_{ri} = P \frac{2 \kappa_i^2}{1 + \kappa_i^2 + \kappa_i^4}\end{split}
\end{equation*}
\paragraph*{Case 3} $  \kappa_i^2 > \left.\frac{1 +
    \sqrt{5}}{2}\right.,$
\begin{equation*}P_{ai} =
P \frac{1}{1 + \kappa_i^2}  \,\,\,\,\,\,  P_{bi} = P \,\,\,\,\,\,
P_{ri} = P \frac{\kappa_i^2}{1 + \kappa_i^2}
\end{equation*}
\end{thm}
\begin{proof}
The problem can be solved analytically using the method of Lagrange
multipliers and applying the Karush-Kuhn Tucker conditions as given
below. The optimization problem for finite $L$ and $\Delta = 0.5$
can be expressed below as, \begin{eqnarray*}  \mbox{maximize     }  && \frac{1}{L} \sum_{i = 1}^L \left(R_{ab}^{(i)} + R_{ba}^{(i)}\right)      \\
         \mbox{subject to     } && \frac{1}{2} C(|h_{ai}|^2P_{ai}) >
                                R_{ab}^{(i)}, \frac{1}{2} C(|h_{bi}|^2P_{ri}) >
                                R_{ab}^{(i)} \\
                                && \frac{1}{2} C(|h_{bi}|^2P_{bi}) >
                                R_{ba}^{(i)}, \frac{1}{2} C(|h_{ai}|^2P_{ri}) >
                                R_{ba}^{(i)} \\
                                &&  \frac{1}{L} \sum_{i=1}^L \left(P_{ai} +   P_{bi} +   P_{ri}\right) \leq 2  P\\
                                && P_{ai}, P_{bi}, P_{ri} \geq 0, i
                                \in \{1,2 \hdots L\}
       \end{eqnarray*}
We next use the high snr approximation to simplify the analysis.
Under the high snr approximation $C(x) \approx \log x$.
 The Lagrangian
can then be expressed as,
\begin{equation*}
\begin{split}
\mathcal{L} = - \frac{1}{L}\sum_{i=1}^L \left(R_{ab}^{(i)} +
R_{ba}^{(i)}\right) \\ + \frac{\lambda}{L} \left(\sum_{i=1}^L
\left(P_{ai} + P_{bi} + P_{ri}\right) - 2P\right) \\ +
\sum_{i=1}^L\mu_1^{(i)} (2 R_{ab}^{(i)} - \log(|h_{ai}|^2P_{ai})) \\
+ \sum_{i=1}^L\mu_2^{(i)} (2 R_{ab}^{(i)} - \log(|h_{bi}|^2P_{ri}))
\\ + \sum_{i=1}^L\mu_3^{(i)}(2 R_{ba}^{(i)} - \log(|h_{bi}|^2P_{bi})) \\ + \sum_{i=1}^L\mu_4^{(i)} (2 R_{ba}^{(i)} -
\log(|h_{ai}|^2P_{ri}))
\end{split}
\end{equation*}
Next taking the partial derivative of $\mathcal{L}$ with respect to
each variable in $\mathcal{L}$ and equating to zero, gives us the
following set of equations.
\begin{equation*} -\frac{1}{L} + 2 \mu_1^{(i)} + 2 \mu_2^{(i)} = 0 ,
 -\frac{1}{L} + 2 \mu_3^{(i)} + 2 \mu_4^{(i)} = 0
\end{equation*}
\begin{equation*} \frac{\lambda}{L} - \frac{\mu_1^{(i)}}{P_{ai}} = 0 ,
 \frac{\lambda}{L} - \frac{\mu_3^{(i)}}{P_{bi}} = 0,  \frac{\lambda}{L} - \frac{\mu_2^{(i)} + \mu_4^{(i)}}{P_{ri}} = 0
\end{equation*}

Solving for $\lambda$ from the above set of equations gives $\lambda
= \frac{1}{2P}$.
 Solving again the above set of equations  for $P_{ai}$, $P_{bi}$ and $P_{ri}$ and together with the
  Karush-Kuhn Tucker(KKT) conditions,
  gives us the  required power allocation  as a
function of $\kappa_i^2$.
\end{proof}

In the above proof,  the parameter $\lambda$  is not a function of
the channel gains. This is so, since we have made the high snr
approximation. This implies that the total power allocated $P_{ai} +
P_{bi} + P_{ri}$, during each coherence time interval remains the
same. However, the  power in the individual nodes will vary based on
the channel gains. From Theorem 1, we can see that in case 1 we have
at low $\kappa_i^2$, $ \kappa_i^2 P_{ai} \approx P_{bi}$. In case 3,
where $\kappa_i^2 >> 1$, $ \kappa_i^2 P_{ai} \approx P_{bi}$. Also
for case 2, where $\kappa_i^2 \approx 1$, $ \kappa_i^2 P_{ai}
\approx P_{bi}$. In other words this implies that $ |h_{ai}|^2
P_{ai} \approx |h_{bi}|^2 P_{bi}$. In the next section, we propose a
scheme that makes use of this property to obtain results close to
the upper bound.

\section{Achievable  Scheme using channel inversion and lattice
coding}\label{sec:Achievable scheme}

In this  section we discuss our achievable scheme based on nested
lattice decoding by Erez and Zamir\cite{erez04}. The proposed scheme
follows closely the lattice coding scheme discussed in our previous
work \cite{narayanan07} and in \cite{nazerallerton07}. The main idea
in \cite{narayanan07}, is that suppose each of the nodes $A$, $B$
and the relay $R$, has the same power constraint (say
$P_{\Lambda}$), then at high signal to noise ratios, a rate close to
the upper bound $C(P_{\Lambda})$ can be exchanged. In other words,
the nested lattice coding approach works best when the receiver
channel signal strengths from the two nodes are the same.

In our proposed scheme, we enforce  $|h_{ai}|^2P_{ai} =
|h_{bi}|^2P_{bi} $ for every $i^{th}$ coherence time interval in the
MAC phase. This matches closely with the observation in section
\ref{sec:upperbound}, where  the power allocation profile at the
nodes satisfies $|h_{ai}|^2P_{ai} \approx |h_{bi}|^2P_{bi}$. This
means that each node uses a coarse lattice of power $P_{\Lambda i}$,
and each node performs a channel inversion at the transmitter, so
that the relay receives equal signal strengths from both the nodes.
For the broadcast phase, to ensure that the nodes $A$ and $B$ can
decode the message from the relay, we enforce that transmit power at
the relay is always larger than the transmit power at the nodes, or
$P_{ri} \geq P_{ai}, P_{bi}$.

First let us explain our achievable scheme in detail for the
$i^{th}$ coherence interval with known channel gains $h_{ai}$,
$h_{bi}$. We first obtain the power allocation profiles $P_{ai},
P_{bi}$ and $P_{ri}$, based on the additional requirements of
$|h_{ai}|^2P_{ai} = |h_{bi}|^2P_{bi} $ and $P_{ri} \geq P_{ai},
P_{bi}$. The allocation is discussed in more detail later in the
section. Let us next define $P_{\Lambda i} := |h_{ai}|^2 P_{ai} =
|h_{bi}|^2 P_{bi}$ and for each coherence time interval $i$, choose
a nested lattice structure having a fine lattice $\Lambda_i^{f}$
with a coarse lattice $\Lambda_i$ nested in it. The second moment
per unit dimension of the coarse lattice is $P_{\Lambda i}/2$. Also
the channel model considered in this problem setup has complex
inputs and complex noise, when compared to the real Gaussian channel
model in \cite{narayanan07}.
 The complex channel provides two degrees of freedom. To take advantage of this
 we can perform nested
 lattice coding separately along the in-phase and the quadrature phase
 components.  In all the vectors discussed below namely $\mathbf{x_{ai}},\mathbf{x_{bi}},\mathbf{t_{ai}},\mathbf{t_{bi}},
 \mathbf{u_{ai}}$ and $\mathbf{u_{bi}}$ are complex vectors and can be
 expressed as the complex sum of their in-phase and quadrature phase
 components. For example $\mathbf{x_{ai}}$ can be expressed as $\Re \{\mathbf{x_{ai}}\} + \jmath \Im \{ \mathbf{x_{ai}} \}.$

  First at each coherence interval $i$ during the MAC phase, the data at the nodes $A$ and $B$ are mapped to lattice
points $\mathbf{t_{ai}}$ and $\mathbf{t_{bi}}$ respectively. Let
$\mathbf{u_{ai}}$ and $\mathbf{u_{bi}}$ be dithers that are
uniformly distributed over the coarse lattice.  The in-phase
component of the dither and the quadrature phase component are
independent of each other and each distributed uniformly in the
coarse lattice $\Lambda_i$, with second moment $P_{\Lambda i}$. An
output $(\mathbf{t_{ai}} - \mathbf{u_{ai}}) \mod \Lambda_i$ is
obtained at node $A$ and $(\mathbf{t_{bi}} - \mathbf{u_{bi}}) \mod
\Lambda_i$ at node $B$. Here $ \mathbf{t_i}\mod \Lambda_i$
represents $ (\Re \{\mathbf{t_i}\} \mod \Lambda_i) + \jmath (\Im
\{\mathbf{t_i}\} \mod \Lambda_i)$

Hence the transmitted vector $\mathbf{x_{ai}}$  at node $A$ is given
by,
\begin{equation}
\mathbf{x_{ai}} = \frac{(\mathbf{t_{ai}} - \mathbf{u_{ai}}) \mod
\Lambda_i}{h_{ai}}
\end{equation}
Here the numerator $ (\mathbf{t_{ai}} - \mathbf{u_{ai}}) \mod
\Lambda_i $ is the sum of the in-phase and quadrature phase
components, expressed as $ \Re(\mathbf{t_{ai}} - \mathbf{u_{ai}})
\mod \Lambda_i + \jmath \Im(\mathbf{t_{ai}} - \mathbf{u_{ai}}) \mod
\Lambda_i$. Hence the second moment of the numerator per unit
dimension is $P_{\Lambda i}/2 + P_{\Lambda i}/2 = P_{\Lambda i}. $
Hence the average transmit power on $\mathbf{x_{ai}}$ is $P_{\Lambda
i}/|h_{ai}|^2 = P_{ai}.$ Thus we meet the average power constraint
of $P_{ai}$ at node $A$. Similarly the transmitted vector
$\mathbf{x_{bi}}$ at node $B$ is given below. This also meets the
average power constraint $P_{bi}$.
\begin{equation}
\mathbf{x_{bi}} = \frac{(\mathbf{t_{bi}} - \mathbf{u_{bi}}) \mod
\Lambda_i}{h_{bi}}
\end{equation}
 The relay receives
\begin{equation}
\mathbf{y_{ri}} = h_{ai}\mathbf{x_{ai}} + h_{bi}\mathbf{x_{bi}} +
\mathbf{z_{ri}}
\end{equation}or
\begin{equation}
\mathbf{y_{ri}} = (\mathbf{t_{ai}} - \mathbf{u_{ai}}) \mod \Lambda_i
+ (\mathbf{t_{bi}} - \mathbf{u_{bi}}) \mod \Lambda_i +
\mathbf{z_{ri}}
\end{equation}

The decoder next forms $(\mathbf{y_{ri}} + \mathbf{u_{ai}} +
\mathbf{u_{bi}}) \mod \Lambda_i $ and performs nested lattice
decoding and decodes to $ \mathbf{t_{ri}} =  (\mathbf{t_{ai}} +
\mathbf{t_{bi}}) \mod \Lambda_i$ with high probability, as long as
the transmission rate from each of the nodes is less than $2 \{
\frac{1}{2} \log (0.5 + P_{\Lambda_i})\}$. The factor 2 is present
because the channel coefficients are complex and we have 2 degrees
of freedom.

The relay next forms $(\mathbf{t_{ri}} - \mathbf{u_{ri}} ) \mod
\Lambda_i$ and during the broadcast phase transmits
\begin{equation}
\mathbf{x_{ri}} =
\sqrt{\frac{P_{ri}}{P_{\Lambda_i}}}\{(\mathbf{t_{ri} -
\mathbf{u_{ri}}) \mod \Lambda_i\}}
\end{equation}
The relays can decode to $\mathbf{t_{ri}}$ as   $|h_{ai}|^2 P_{ri},
|h_{bi}|^2 P_{ri} \geq P_{\Lambda_i}$, since $P_{ri} \geq P_{ai},
P_{bi}$. Hence effectively a rate of $\frac{1}{2} \log (0.5 +
P_{\Lambda i})$ can be achieved by the nodes.

 Also
define $D(x) :=  {u.c.e}\{\log(0.5 + x), 0.5 \log (1 + 2x)\}, x \geq
0$. Here $u.c.e$ denotes the upper concave envelope of the two
functions. Hence the optimization problem can be expressed as
follows with a few more constraints added.
  \begin{eqnarray}
         \mbox{maximize }  &&   \frac{1}{L}\sum_{i=1}^{L} \min \{ \Delta D(|h_{ai}|^2P_{ai}), (1-\Delta) D(|h_{bi}|^2P_{ri})
         \}\nonumber\\ && + \frac{1}{L}\sum_{i=1}^{L} \min \{ \Delta D(|h_{bi}|^2P_{bi}), (1-\Delta) D(|h_{ai}|^2P_{ri}) \}  \nonumber \\
         \mbox{subject to } && \frac{\Delta}{L} \sum_{i=1}^{L} P_{ai} + \Delta \sum_{i=1}^{L} P_{bi} + \frac{(1 -
                                    \Delta)}{L} \sum_{i = 1}^{L} P_{ri} \leq  P,\\
                                && |h_{ai}|^2P_{ai} =
                                |h_{bi}|^2P_{bi}, \\
                                && P_{ri} \geq P_{ai},\label{eqn:P_r geq P_a} \\
                                && P_{ri} \geq P_{bi},\label{eqn:P_r geq P_b} \\
                                && P_{ai}, P_{bi}, P_{ri} \geq 0, i
                                \in {1,2 \hdots L} \nonumber
       \end{eqnarray}
\begin{figure}[thpb]
      \centering
      \includegraphics[scale=0.4]{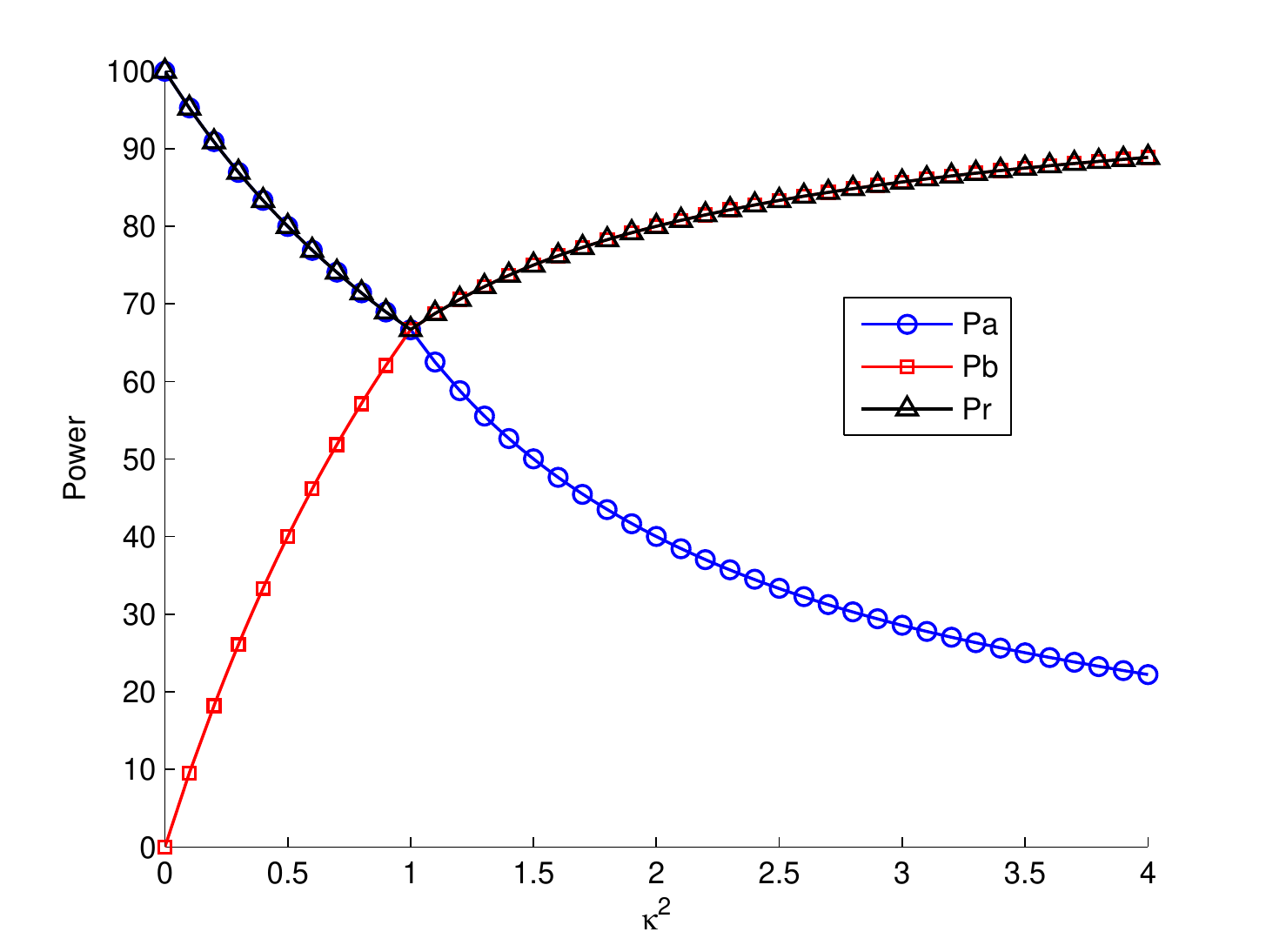}
      \caption{Power allocation as a function of $\kappa^2$ for our proposed scheme}
      \label{fig:chinv_bd_power_profile}
   \end{figure}
The above optimization problem is solved for the case for $\Delta =
0.5$ under the high snr assumption with $D(x)$ approximated by $\log
(x)$. The next theorem gives us the power allocation profile at the
different nodes for different values of $\kappa_i^2 :=
|h_{ai}|^2/|h_{bi}|^2$.
\begin{thm}
For finite  $L$ with $\Delta = 0.5$ and under the high snr
approximation, the optimal power allocation for the  achievable
scheme for different ranges of $\kappa_i^2 := |h_{ai}|^2/|h_{bi}|^2$
is given by,
\paragraph*{Case 1} $ 0 < \kappa_i^2 <= 1,$
\begin{equation*}P_{ai} = P \frac{2}{2 + \kappa_i^2 }  \,\,\,\,\,\,  P_{bi} = P \frac{2\kappa_i^2}{2 + \kappa_i^2}
\,\,\,\,\,\, P_{ri} = P \frac{2}{2 + \kappa_i^2}
\end{equation*}
\paragraph*{Case 2} $  \kappa_i^2 > 1,$
\begin{equation*}P_{ai} =
P \frac{2}{1 + 2 \kappa_i^2}  \,\,\,\,\,\,  P_{bi} = P \frac{2
\kappa_i^2}{1 + 2 \kappa_i^2} \,\,\,\,\,\, P_{ri} = P \frac{2
\kappa_i^2}{1 + 2 \kappa_i^2}
\end{equation*}
\end{thm}
\begin{proof}
The problem can be solved analytically using the method of Lagrange
multipliers and using the Karush-Kuhn Tucker conditions following
along the same lines as in the proof of Theorem 1.
\end{proof}

\section{Comparison of the upper bound and the achievable
schemes}\label{sec:Comparison}


%
%
%

Next we state the main theorem of this paper that compares the upper
bound and the achievable scheme for known channel state information
in $L$ coherence intervals.
 \begin{figure}[thpb]
      \centering
      \includegraphics[scale=0.4]{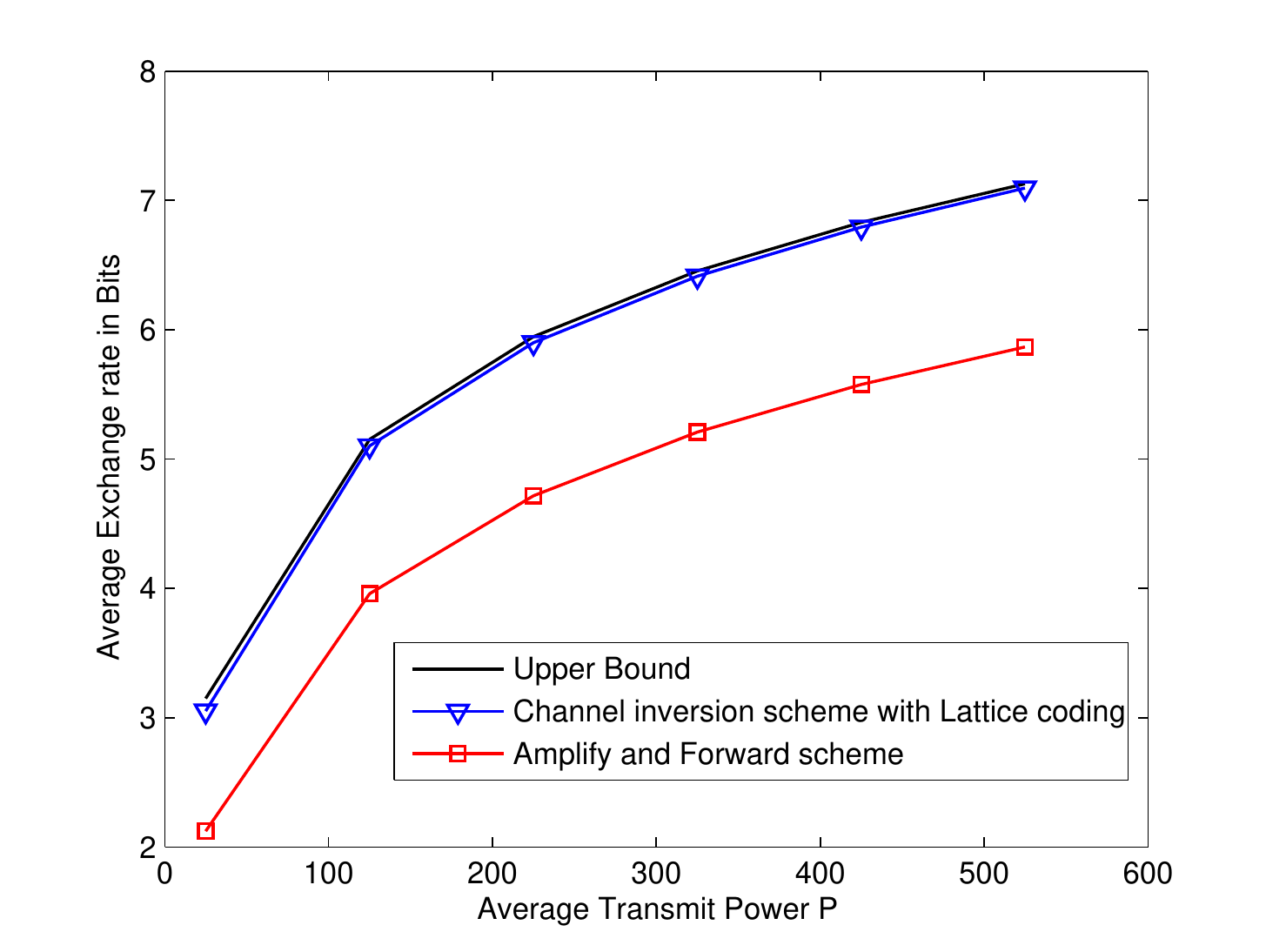}
      \caption{Comparison of bounds for $\Delta = 0.5$}
      \label{fig:fading_comparison}
   \end{figure}
\begin{thm}
For general  $L$ with $\Delta = 0.5$, and under the high snr
approximation, the achievable rate using the channel inversion
scheme with lattice decoding suffers at most a constant $\eta =
0.08972$ bits per complex channel use from the upper bound.
\end{thm}
\begin{proof}
We  compare analytically the results of Theorem 1 and Theorem 2
under the high snr approximation. For different values of $\kappa_i$
we compare the exchange rates and we can easily show that the
channel inversion scheme suffers at max $0.08972$ bits per complex
channel use from the upper bound.
\end{proof}

 Theorem 3, hence captures the loss due to adding the additional constraints to the
 upper bound, and the loss is found to be really small.  Fig.\ref{fig:fading_comparison} shows some
 results obtained by numerical solving the
 optimization problems without the high snr approximation.
  Here we have $L =100$, and the channel coefficients are taken from a Rayleigh distribution, and
 they  have unit variance. The channel coefficients are fixed and we evaluate the average rate per channel use,
 each  for the upper bound,  the lattice based scheme and also the amplify forward
 scheme.

\section{Practical issues with power allocation
design}\label{Practical Issues}

In the previous sections, we discussed techniques to compute the
optimal power allocation for a given $\mathbf{h_a}, \mathbf{h_b}$
and $\mathbf{h_r}$. However, in practice we are interested in
maximizing the average exchange capacity, i.e., the problem is to
     \begin{eqnarray*}
         \mbox{maximize     }  &&   \mathbb{E}_{h_a,h_b}\left[ \min \{ \Delta C(|h_{a}|^2P_{a}), (1-\Delta)
         C(|h_{b}|^2P_{r})\}\right.
         \\ && \left.              + \min \{ \Delta C(|h_{b}|^2P_{b}), (1-\Delta) C(|h_{a}|^2P_{r}) \}\right] \\
         \mbox{subject to     } && \mathbb{E}_{h_a,h_b} \left[ \Delta  P_{a}  + \Delta  P_{b} + (1 -
\Delta)  P_{r}  \right] \leq P\\
                                && P_{a}, P_{b}, P_{r} \geq 0.
       \end{eqnarray*}
Due to the ergodic nature of the channel, the above optimization
problem is identical to the one in (\ref{eqn:upperbound}) when $L
\rightarrow \infty$. The result in Fig.~\ref{fig:fading_comparison}
have been obtained by solving the optimization problem for one
realization of $\mathbf{h_a}, \mathbf{h_b}$ and $\mathbf{h_r}$ but
with $L=100$. Note that while computing the optimal power allocation
policy requires us to use a large value of $L$ and optimize the
policy, once this policy is fixed, the actual transmit power is
chosen based only on the instantaneous channel realization.



%

\section{Conclusion}\label{sec:Conclusion}
In this paper we studied the bi-directional relay problem in which
the channels between the nodes and the relays were assumed to have
complex inputs with complex fading coefficients. We studied power
allocation policies at the nodes for a two phase transmission scheme
under the sum transmit power constraint over all nodes. For $\Delta
= 0.5$, where each phase uses the channel exactly half the time, we
obtained an upper bound on the exchange capacity as a solution to a
convex optimization problem. We proposed a scheme using nested
lattice encoding with the transmit power chosen to be inversely
proportional to the channel gains. We obtained analytical solutions
for the exchange capacity under the high snr approximation and
showed that our proposed scheme can obtain a rate which is at most
$0.09$ bits away from the upper bound. For $\Delta \neq 0.5$, we
were unable to obtain a good performance using a simple channel
inversion power allocation policy.
 However, it can be shown that using lattice codes with asymmetric rates \cite{nam08} at the nodes,
 the upper bound can be achieved at high snrs.


\begin{thebibliography}{1}

\bibitem{SangJoonKim08}
 S.~J.~Kim, N.~Devroye, P.~Mitran and V.~Tarokh,  ``Achievable rate regions for bi-directional relaying'',  arxiv.org 2008.

%
%
%
%
%
%
%
%
%
%
%
%
%
\bibitem{erez04}
U.~Erez and R.~Zamir, ``Achieving $\frac{1}{2} \log(1 +
\textrm{SNR})$ on the AWGN channel with lattice encoding and
decoding,'' {\em {IEEE} Tran. Info. Theory}, vol. 50, pp. 2293–2314,
October 2004.

%
%
%
%
\bibitem{narayanan07}
K.~R.~Narayanan, M.~P.~Wilson and A.~Sprintson,``Joint Physical
Layer Coding and Network Coding for Bi-Directional Relaying'',
{\em{45th Annual Allerton Conference on Communication, Control and
Computing}}, September 2007.

\bibitem{nazerallerton07}
B.~Nazer and M.~Gastpar, ``Lattice Coding Increases Multicast Rates
for Gaussian Multiple-Access Networks'', {\em{45th Annual Allerton
Conference on Communication, Control and Computing}}, September
2007.

\bibitem{baik08}
I.~J.~Baik and S.~Y.~Chung, ``Network coding for two-way relay
channels using lattices'', {\em{ Proc. IEEE International Conference
on Communications}}, Beijing, China, May 2008.

\bibitem{nam08}
W.~Nam, S.~Y.~Chung and Y.~H.~Lee, `` Capacity bounds for two-way
relay channels'', {\em{Proc. IEEE International Zurich Seminar on
Communications}}, Zurich, Switzerland, March 2008.


%
%
%
%
%
%
%
%
%
%
%
%
%
%
%
%
%
%
%
%
%
%
%
%
%
%
%
%
%
%
%
%
%
%
%
%
%
%


\end{thebibliography}
\end{document}